\documentclass{optica-article}

\journal{ao}

\articletype{Research Article}

\usepackage{lineno}

\usepackage{tabularx} 
\usepackage{amsmath}  
\usepackage{mathtools} 
\usepackage{physics}
\usepackage{graphicx} 
\usepackage{float} 
\usepackage{mathtools} 
\usepackage{color}  

\definecolor{mscolor}{rgb}{0,0.5,0.5}
\definecolor{phcolor}{rgb}{0.5,0,0.5}
\definecolor{tgcolor}{rgb}{0.5,0,0.5}
\definecolor{eocolor}{rgb}{0.5,0.5,0}

\begin{document}

\title{Multi-scale architecture for fast optical addressing and control of large scale qubit arrays}

\author{T. M. Graham, E. Oh, and M. Saffman}

\address{Department of Physics, 1150 University Ave., University of Wisconsin-Madison, Madison, WI, 53706, USA
}

\email{msaffman@wisc.edu}

\begin{abstract}
We present a technique for rapid site-selective control of the quantum state of particles in a large array using a combination of a fast deflector (e.g. an acousto-optic deflector) and a relatively slow spatial light modulator. The use of spatial light modulators for site-selective quantum state manipulation has been limited due to slow transition times preventing rapid, consecutive quantum gates. By partitioning the spatial light modulator into multiple segments, and using a fast deflector to transition between them, it is possible to substantially reduce the average time increment between scanner transitions by increasing the number of gates that can be performed for a single spatial light modulator full frame setting.  We analyze the performance of this device in two different configurations: in configuration 1, each segment of the spatial light modulator addresses the full qubit array; in configuration 2, each segment of the spatial light modulator addresses a sub-array and an additional fast deflector positions that sub-array with respect to the full qubit array. With these hybrid scanners we calculate  qubit addressing rates that are tens to hundreds of times faster than using an SLM alone.  
\end{abstract}

\section{Introduction}

The last decade has seen the demonstration of the first small and medium scale quantum computing devices. The demanding requirements for quantum computing have led to several different competing platforms using different physical qubit technologies. One promising approach encodes qubits in internal states of neutral atoms, which are trapped in an optical lattice contained in a vacuum cell \cite{Saffman2016}. Such atomic lattices promise long qubit coherence times, controllable interactions, and identical qubits; however, perhaps the most attractive feature of an atom-based quantum computer is the ability to scale to large numbers of qubits. This trait arises from flexibility and ease of construction of large optical lattices using the mature technologies of high-power lasers and various light patterning devices such as holograms and spatial light modulators (SLMs). Optical lattices composed of $>1000$ traps have already been demonstrated\cite{DKim2019,Huft2022} and arrays with more than 300  deterministically positioned atoms have been achieved\cite{Schymik2022}. The size of such arrays is only limited by the available optical power and the array trapping optics. Lattices with $>10,000$ trapping sites, based on existing technology, can realistically be anticipated in the next few years. 

However, rapidly performing quantum gates on such a large number of qubits 
remains a challenge. Standard techniques for performing multi-qubit gates (or multiple parallel single-qubit gates) on neutral atom qubits require the ability to optically address targeted qubits with focused lasers without illuminating non-targeted qubits.  Techniques to perform multi-qubit addressing on small-scale optical lattice quantum computers\cite{Graham2022} are not efficiently scalable to larger arrays and/or multi-qubit gates that require simultaneous illumination of multiple qubits. Available technologies for rapidly scanning optical beams across multiple spatial locations have been reviewed in 
\cite{Bechtold2013,Romer2014}. As pointed out in these references, there are technologies such as piezoelectric devices, galvanometer mirrors, and liquid crystal or digital micro mirror device (DMD) spatial light modulators that can flexibly address many spatial locations but are relatively slow. Technologies that are very fast, such as acousto-optic and electro-optic beam scanners, cannot arbitrarily address many spatial locations. We reveal here a new architecture that combines fast and slow devices to achieve rapid addressing of a large number of spatial locations. 

Scalable control of large qubit arrays is challenging due  to the competing requirements of addressing multiple trap sites simultaneously and being quickly reconfigurable while  allowing many gates to be performed in the qubit coherence time.  The former requirement can be achieved using SLMs. These devices allow a spatially varying amplitude and/or phase mask to be applied to an incident light field.  When used as a hologram, an arbitrary illumination pattern can be generated on the qubit array, allowing the addressing of multiple qubit sites with high diffraction efficiency. These devices have the additional benefit of also being able to shape optical beams to allow more uniform illumination \cite{Ebadi2021} and correct some optical aberrations in the beam path \cite{Bowman2010, Nogrette2014,Shih2021, Ebadi2021}.  Unfortunately, the switching speed of such spatial light modulators is slow, typically 1 kHz or less for liquid crystal-based SLMs. DMDs may be modulated at up to 20 kHz but only provide binary amplitude modulation, which limits the diffraction efficiency to $\leq 10\%$\cite{Turtaev2017}. 

In contrast to SLM qubit addressing, fast deflectors such as electro-optic deflectors (EOD) and acousto-optic deflectors (AODs) provide efficient, quickly reconfigurable qubit addressing, but are limited in their ability to address multiple sites simultaneously\cite{Schmidt-Kaler2003b,Graham2022}.  EODs deflect an incoming light beam using  either Kerr or Pockels interactions. While very efficient, EODs are currently bulkier, less commercially available, and have a smaller scan range than AODs. For this reason, we will analyze the performance of this scanner in the context of using AODs rather than EODs; however much of this analysis can be extended to any beam deflector technology.  In an AOD, a transducer creates a traveling acoustic wave in a crystal; the spatially varying index of refraction caused by this wave acts as a traveling bulk diffraction grating on an incoming light beam. This traveling wave both diffracts the light and gives the first-order diffracted beam a frequency shift equal to the acoustic wave frequency. By adjusting the acoustic wave driving frequency, the angle of the output diffracted beam can be controlled in one dimension.  Using two perpendicularly oriented AODs, an input beam can be steered in two dimensions. To address multiple sites, an AOD is driven with multiple frequencies; however, this addressing scheme cannot be used to simultaneously address arbitrary sites in a qubit array. When steering a single optical beam with a 2D AOD scanner, two frequencies are needed, a horizontal frequency  $f_{1,x}$ and a vertical frequency $f_{1,y}$. However, driving horizontal and vertical AODs each with two different frequencies, $f_{1,x}$, $f_{2,x}$, $f_{1,y}$, $f_{2,y}$, will NOT result in two diffracted beams (one corresponding to the first $x,y$ frequency pair and one corresponding to the second frequency pair) because nothing prevents diffraction at cross-indexed frequency combinations. Driving with two horizontal and two vertical frequencies will instead result in four diffracted beams corresponding to frequency combinations: $(f_{1,x},f_{1,y})$, $(f_{1,x},f_{2,y})$, $(f_{2,x},f_{1,y})$, and $(f_{2,x},f_{2,y})$. More generally, if there are $N_l$ frequencies driving the horizontal AOD and $M_l$ frequencies driving the vertical AOD, then an $N_l \times M_l$ grid of diffracted beam spots will be output. This limitation significantly limits the utility of crossed AODs for performing arbitrary addressing of sites in a qubit array.

In addition to limitations in the addressing pattern, frequency shifts can cause complications in addressing multiple qubit sites. As mentioned above, when a beam is diffracted by an AOD, the diffracted light is shifted in frequency by the AOD driving frequency. Since addressing multiple array sites requires driving at least one axis of a 2D AOD with multiple frequencies, the output beams will not all have the same frequency. This is problematic when the optical beams are being used to drive resonant transitions, since the detuning from the transition will change over the beam spot pattern. In addition, a small amount of crosstalk between diffracted beams from optical aberrations or scattering can have a large time-varying effect on the intensity of the light addressing the qubits (see \cite{Graham2022}).

While both SLMs and AODs individually are limited in their ability to address qubits in a two-dimensional array, we introduce here an architecture that uses both devices in conjunction, thereby enabling large-scale, rapidly reconfigurable, and abitrary addressing of large qubit arrays.   The techniques described here are applicable to a variety of qubit technologies that are controlled with light beams. These include qubits implemented in neutral atoms, molecules, trapped ions, and semiconductor quantum dots.

\begin{figure}[!t]
\includegraphics[width=5in]{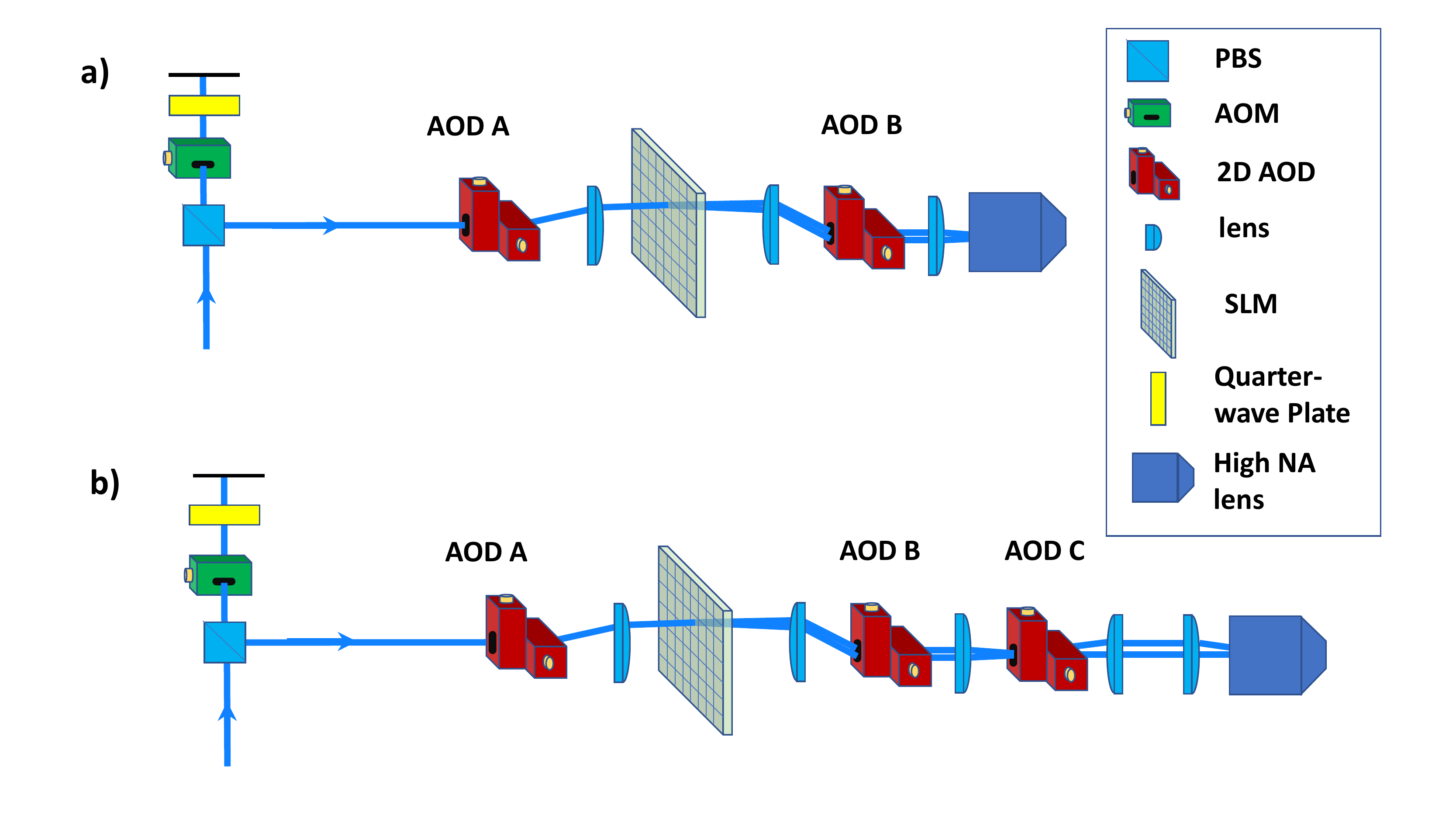}
\caption{Two configurations which combine acousto-optic deflectors (AOD) and spatial light modulators. \textbf{a)} Configuration 1. A 2D AOD (AOD A) is used to select different segments of a spatial light modulator (SLM). Each segment of the spatial light modulator contains a different hologram corresponding to an illumination pattern in the qubit array. A second AOD (AOD B) is used to correct momentum shifts imparted by the first AOD. An upstream double-pass acousto-optic modulator (AOM) is used for fast light beam modulation and frequency control and compensation for other acousto-optic devices. This double pass setup uses a polarizing beam splitter (PBS) and quarter waveplate (QWP) to direct the light beam towards the AODs after passing through the AOM twice. Fast transitions between the SLM partitions using the AODs allow an effective increase in the average SLM reconfiguration rate.  \textbf{b)} Configuration 2. The addition of a third AOD (AOD C) allows fast positioning of the image created by the SLM relative to the qubit array. By including all required addressing configurations in the SLM partition, this configuration can address a large array with fast scanner transition times. Note that  SLMs in transmission  mode are  used in the diagrams; however, modifications for a reflection mode SLM are straight forward, and only require an adjustment of beam path.
\label{fig.configurations}
}
\end{figure}

\section{Hybrid Scanner Description and Modeling}

 We propose to combine the fast deflection capabilities of AODs with the arbitrary addressing and beam shaping capabilities of an SLM in a single optical layout (see Fig. \ref{fig.configurations}).  This hybrid scanner uses an AOD to direct a beam at different regions of the SLM; each of these different regions contains a phase hologram which produces a distinct addressing pattern in the plane where the qubit array is located. Since the AOD allows rapid switching between different regions of the SLM, rapid transitions between addressing patterns are possible. This rapid switching allows faster transitions between quantum gate sets than using only a non-segmented SLM which has several benefits for the operation of a  quantum computer.  More rapid transitions reduce the total time needed for a quantum computation allowing a faster overall computation improving the overall computer performance, since the quantum state degrades over time  due to finite $T_1$ and $T_2$ coherence times.   Fast switching times also enable certain quantum gate procedures that are not possible with  slow scanner transition times. For example, some previously demonstrated 2-qubit controlled phase gates on neutral atom qubits relied on very fast scanner pointing  transition times since some population was left in fragile, highly excited Rydberg states between scanner transitions \cite{Graham2019}. For these gates $<1~\mu\rm s$ scanner transition times were needed to prevent significant Rydberg decay and dephasing, more than 3 orders of magnitude faster than an SLM.  Details of the optical layout and SLM segmentation strategy greatly affect the scanner performance.  Below we detail two possible strategies and provide estimations of the scanner performance in these configurations. The first of these configurations allows global, arbitrary addressing of the entire array with a moderate speed improvement. The second configuration provides arbitrary addressing on a movable sub-array and allows a much more dramatic speed increase at the cost of losing arbitrary addressing over the entire qubit array.

\begin{table}[!t]
\caption{\label{tab.displacement} If AOD B is not used to correct the k-vector shift imparted by AOD A, then the beam position will be displaced on the high NA lens (see Fig. \ref{fig.configurations}). The maximum displacement, $D_{\rm max}$, will depend on the number of SLM partitions, the ratio of the SLM partition size to the beam waist, $q_{\rm SLM}$, and the waist of the beam at the high NA lens, $w_{\rm lens}$ (see eq. (\ref{eq.displacement})). The waist $w_{\rm lens}$ is constrained by the target waist size at the atoms, $w_{\rm a}$, the focal length of the high NA lens, and the wavelength of the light. Below, we calculate $D_{\rm max}$ for a number of configurations where the wavelength is 459 nm, the focal length of the high NA lens is 23 mm, and $q_{\rm SLM}=5$. For this configuration, we observe that AOD B is needed to avoid beam clipping and coma even for a small number of  SLM partitions and moderate $w_{\rm a}$. Note that a lens with NA$=0.7$ and the given focal length has a diameter of 45 mm.}
\centering
\begin{tabular}{ |c|c|c|c|  }
 \hline
 \multicolumn{4}{|c|}{Beam displacement at high NA lens} \\
 \hline
 SLM partitions & $w_{\rm a}~(\mu\rm m)$ & $w_{\rm lens}~\rm (mm)$ & $D_{\rm max}~\rm (mm)$\\
\hline
  $2 \times 2$ & $1.5$  & $2.25$ & $16$ \\
  $2 \times 2$ & $3 $  & $1.12$ & $8$ \\
  $15 \times 15$ & $1.5 $ & $2.25$ & $119$ \\
  $15 \times 15$ & $3 $ & $1.12$& $59$  \\
\hline
\end{tabular}
\end{table}

\subsection{Configuration 1}

\begin{figure}[!t]
\includegraphics[width=5in]{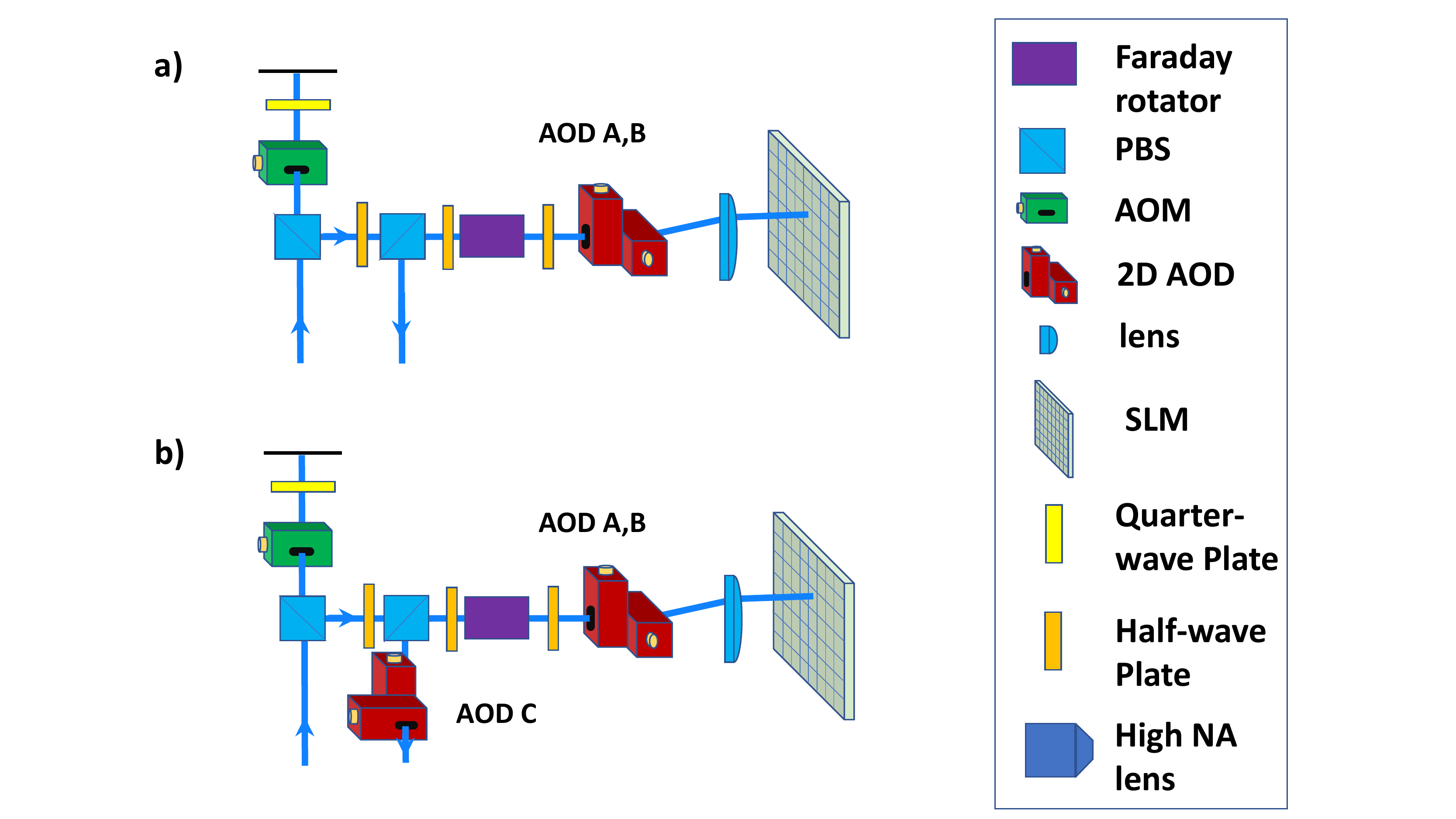}
\caption{Simplified versions of \textbf{a)} Configuration 1 and \textbf{b)} Configuration 2 using a reflective spatial light modulator. The addition of a Faraday rotator, half-wave plates, and a polarizing beam splitter allows the roles of acousto-optic deflectors (AODs) A and B to be combined into a single 2D AOD (AOD A,B). 
\label{fig.configurations_simplified}
}
\end{figure}

The first hybrid AOD and SLM scanner configuration detailed in this paper is composed of an SLM sandwiched by two 2D AODs (see Fig. \ref{fig.configurations}).  The first 2D AOD (AOD A) directs the beam to one of several previously defined regions of the SLM. The SLM regions each impart a spatially-varying phase on the optical beam, shaping the beam to address the targeted array sites. The second 2D AOD (AOD B) then compensates the deflection introduced by AOD A. The correction provided by AOD B prevents clipping on the beam stop of the high numerical aperture lens, which focuses the optical beam on the target array sites.  If we assume Gaussian optics, without AOD B, the maximum displacement, $D_{\rm max}$, at the high numerical aperture lens will be equal to 
\begin{equation}
    D_{\rm max}=q_{\rm SLM} w_{\rm lens}\frac{\sqrt{N_x^2+N_y^2}}{2},
    \label{eq.displacement}
\end{equation}
where $q_{\rm SLM}$ is the ratio of the SLM region size to the beam waist on the SLM and $N_x$ ($N_y$) is the number of horizontal (vertical) SLM partitions. The size of the waist at the atoms is determined by $w_{\rm lens}$, the wavelength of the light, and the focal length of the high NA lens. Table \ref{tab.displacement} lists maximum displacement values for various SLM partition and focusing conditions.  The $k$-vector correction also improves beam focus in the qubit plane by reducing coma and polarization problems due to the light's $k$-vector being misaligned with respect to the optical axis. The other lenses in the system transform the beam between image and Fourier conjugate planes of the qubit array. When using a reflective SLM (or a transmission hologram with a reflector), it is possible to simplify this setup to use only a single 2D AOD with the addition of a Faraday rotator, half-wave plates, and polarizing beam splitter (see Fig. \ref{fig.configurations_simplified}). For clarity, we will describe the operation of this scanner using the layout in Fig.  \ref{fig.configurations}, but similar analysis will also describe the scanners in Fig.  \ref{fig.configurations_simplified}.

\begin{figure}[!t]
\includegraphics[width=5in]{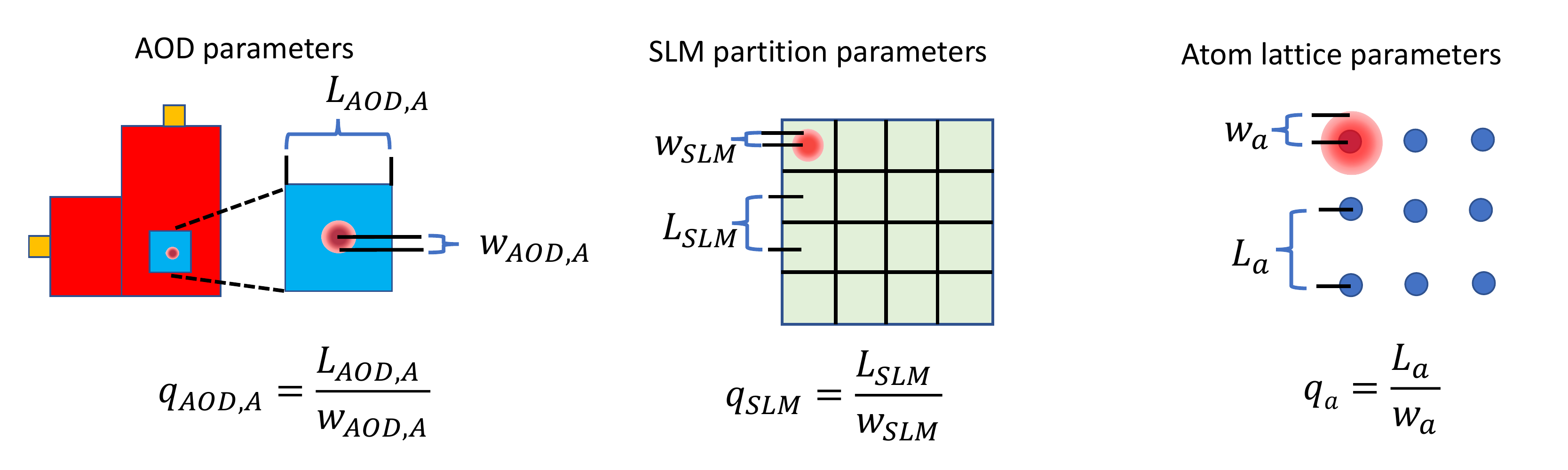}
\caption{\small A summary of experimental parameters used in the text. Gaussian beam waist parameters, $w_{\rm AOD,A}$, $w_{\rm SLM}$, and $w_{\rm a}$ correspond to the $1/e^2$ radius of the Gaussian beam intensity profile in the acousto-optic deflectors (AODs), the spatial light modulator (SLM), and the qubit array, respectively. Note that beam waists in AOD A and AOD B will be identical, but the waist of the beam in AOD C in configuration 2 will be different ($w_{AOD,C}$) and will depend on the lenses used. The length parameters correspond to the AOD active aperture size ($L_{\rm AOD}$), the size of each SLM patch ($L_{\rm SLM}$), and the distance between adjacent qubits in the array plane ($L_{\rm a}$).  Finally, $q_{\rm AOD, A}$, $q_{\rm SLM}$, and $q_{\rm a}$ are ratios of the respective length ($L$) to the corresponding beam waist ($w$). 
\label{fig.scanner_parameters}
}
\end{figure}

To calculate the performance of this hybrid AOD $+$ SLM scanner, the number of addressable array sites and the number of SLM sub-regions  must be considered. These two quantities are jointly constrained by the AOD time-bandwidth product, the SLM pixel resolution, and desired beam spot size in the qubit array image plane. The parameters used in the analysis are summarized in Fig. \ref{fig.scanner_parameters}.
Assuming a Gaussian beam input into AOD A, the maximum number of hologram segments which the SLM can be divided into is limited by the number of resolvable spots ($N_r$) of AOD A
\begin{equation}
    N_r = \frac{\pi}{q_{\rm AOD, A} q_{\rm SLM}}TBW.
    \label{eq.1}
\end{equation}
Here $q_{\rm AOD, A}=L_{\rm AOD}/w_{\rm AOD,A}$ is the ratio of the length of the active aperture of AOD A ($L_{\rm AOD}$) to the Gaussian beam waist in the AOD ($w_{\rm AOD,A}$), $q_{\rm SLM}=L_{\rm SLM}/w_{\rm SLM}$ is the ratio of the SLM segment size $L_{\rm SLM}$ to the beam waist size on the SLM ($w_{\rm SLM}$), and $TBW=T_{\rm{AOD}} \Delta f$ is the time-bandwidth product specification of the AOD, which is the time it takes for an acoustic wave to propagate across the AOD's active aperture ($T_{\rm{AOD}}$) multiplied by the frequency bandwidth of the AOD ($\Delta f$).  

\begin{figure}[!ht]
\centering
\includegraphics[width=4in]{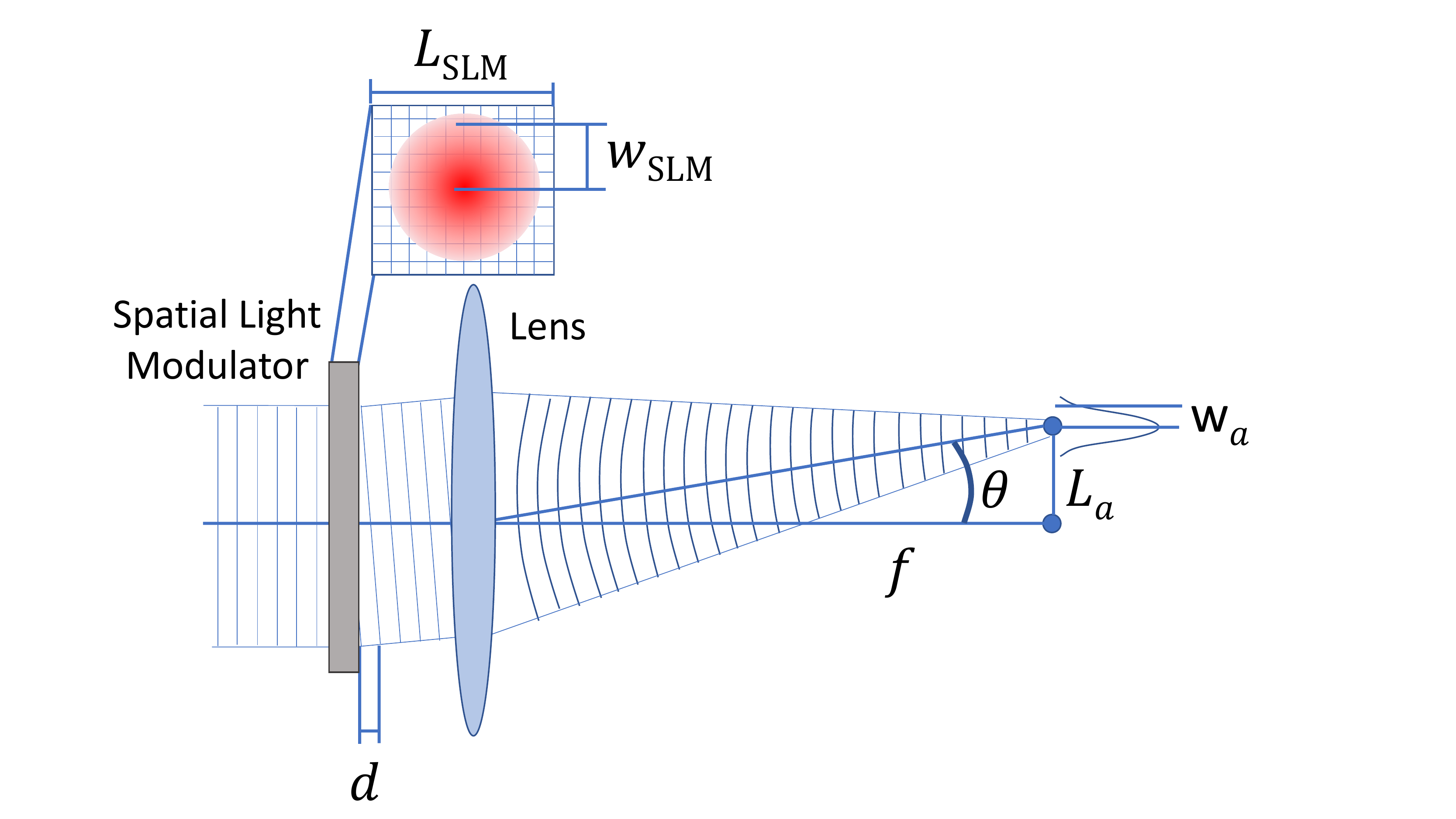}
\caption{A simplified spatial light modulator (SLM) scanner imaging system. An SLM patch applies a phase tilt to an incoming collimated Gaussian beam with waist $w_{\rm SLM}$. The beam diffracts at an angle $\theta$ relative to the incoming beam based on the effective path length difference $d$ that is applied over the SLM patch length, $L_{\rm SLM}$. A lens with focal length $f$ focuses the light onto the qubit array resulting in a Gaussian beam spot with a waist $w_{\rm a}$ that is displaced by $L_{\rm a}$ from the optical axis.  
\label{fig.slm_res}
}
\end{figure}

Each segment of the  SLM will have $n_x \times n_y$ pixels. For this analysis, we use only $+1$ diffraction orders of the SLM to reduce crosstalk onto unwanted target sites due to imperfect diffraction. To calculate the qubit array dimension that each SLM section can address, we need to determine how much displacement the SLM patch can generate in an image plane of  the qubit array relative to the beam waist in that same plane. For this analysis, we can ignore the AODs and consider a simplified setup which just involves illuminating the SLM patch with a collimated Gaussian beam, then focusing the diffracted light on the qubit array plane with a lens (see Fig. \ref{fig.slm_res}). We consider a beam  with  waist $w_{\rm SLM}$ incident on an SLM patch with length $L_{\rm SLM}$. The SLM then applies a 1D phase tilt in the $x$-plane changing the beams momentum.  Focusing the beam with a lens gives a Gaussian spot at a focal length $f$ away from the lens. This spot will be shifted from the optical axis by an amount which depends on the focal length of the lens and the wavefront tilt applied with the SLM. The angle (see Fig. \ref{fig.slm_res})
\begin{equation}
\theta \approx \frac{d}{L_{\rm SLM}}   
\end{equation}
is the angular change that the SLM must induce to the wavefront in order to shift the focus between two adjacent qubit sites in the image plane of the lens.  Since the beam propagates a focal distance from the lens to reach the qubit array, after this propagation, the beam will be displaced by a distance of 
\begin{equation}
L_{\rm a} \approx \theta f = \frac{f d}{L_{\rm SLM}},
\label{eq.La}
\end{equation}
where $L_{\rm a}$ is the distance between neighboring qubit sites. If the SLM applies $v$ full wave fringes across the SLM patch (i.e. the phase linearly increases from $0$ to $v \times 2 \pi$), then we can identify the total phase shift over the aperture as
\begin{equation}
\phi_{\rm tot}= \frac{2 \pi d}{\lambda} = 2 \pi v,
\label{eq.phi}
\end{equation}
where $\lambda$ is the wavelength of the light.  We can now relate $L_{\rm a}$ to the parameters of the Gaussian beam incident on the SLM, the SLM phase mask, and the focal length of the following lens by solving for $d$ in eq.  (\ref{eq.phi}) and substituting into eq. (\ref{eq.La}) yielding
\begin{equation}
L_{\rm a} \approx \frac{v \lambda f}{L_{\rm SLM}}.
\label{eq.La_final}
\end{equation}

For single-site addressing in the qubit array, we define an inter-site spacing ($L_{\rm a}$) to addressing beam waist ($w_{\rm a}$) ratio as $q_{\rm a}=\frac{L_{\rm a}}{w_{\rm a}}$. We can relate $w_{\rm a}$ to the beam parameters just after the SLM and lens system using well-known formulae for the radius of a Gaussian beam assuming the beam waist is in the plane of the qubit array
\begin{equation}
    w_{\rm SLM}=w(-f)=w_{\rm a} \sqrt{1+\left(\frac{-f}{z_R} \right)^2},
\end{equation}
where $z_R=\frac{\pi w_{\rm a}^2 }{\lambda}$ is the Rayleigh range of the beam at the qubit array. Assuming $\frac{f}{z_R} \gg 1$, we can solve for $w_a$
\begin{equation}
    w_{\rm a} \approx  \frac{f \lambda }{\pi w_{\rm SLM}} = \frac{f \lambda q_{\rm SLM}}{\pi L_{\rm SLM}}.
\label{eq.wa_final}
\end{equation}
Now we can relate $q_{\rm a}$ to $v$ using  eqs. (\ref{eq.La_final},\ref{eq.wa_final}) to get
\begin{equation}
    q_{\rm a} \approx \frac{v \lambda f}{L_{\rm SLM}} \frac{\pi L_{\rm SLM}}{\lambda f q_{\rm SLM}} = \frac{\pi v}{q_{\rm SLM}}.
\end{equation}
The number of full $2 \pi$ fringes the SLM needs to apply to the incident beam to move one site in the image plane is then
\begin{equation}
    v \approx \frac{q_a q_{\rm SLM}}{ \pi}.
    \label{eq.v}
\end{equation}
With $n_x$ pixels, we can apply a maximum number of fringes of $v_{\rm max} = \frac{n_x}{2}$ by alternating between $0$ and $\pi$ phase shifts. Dividing this by the number of fringes needed per site, we find that the maximum number of sites that can be addressed along one dimension is 
\begin{equation} 
N_{q,\rm max}=\frac{\pi n_x}{2 q_{\rm a} q_{\rm SLM}}.
\label{eq.Nqmax}
\end{equation}
We can apply the same analysis to the other spatial dimension, yielding, a maximum addressable array size of 
$\frac{\pi n_x}{2 q_{\rm a} q_{\rm SLM}} \times \frac{\pi n_y}{2 q_{\rm a} q_{\rm SLM}}$ sites that can be addressed in the array plane.

\begin{figure}[!t]
\centering
\includegraphics[width=4in]{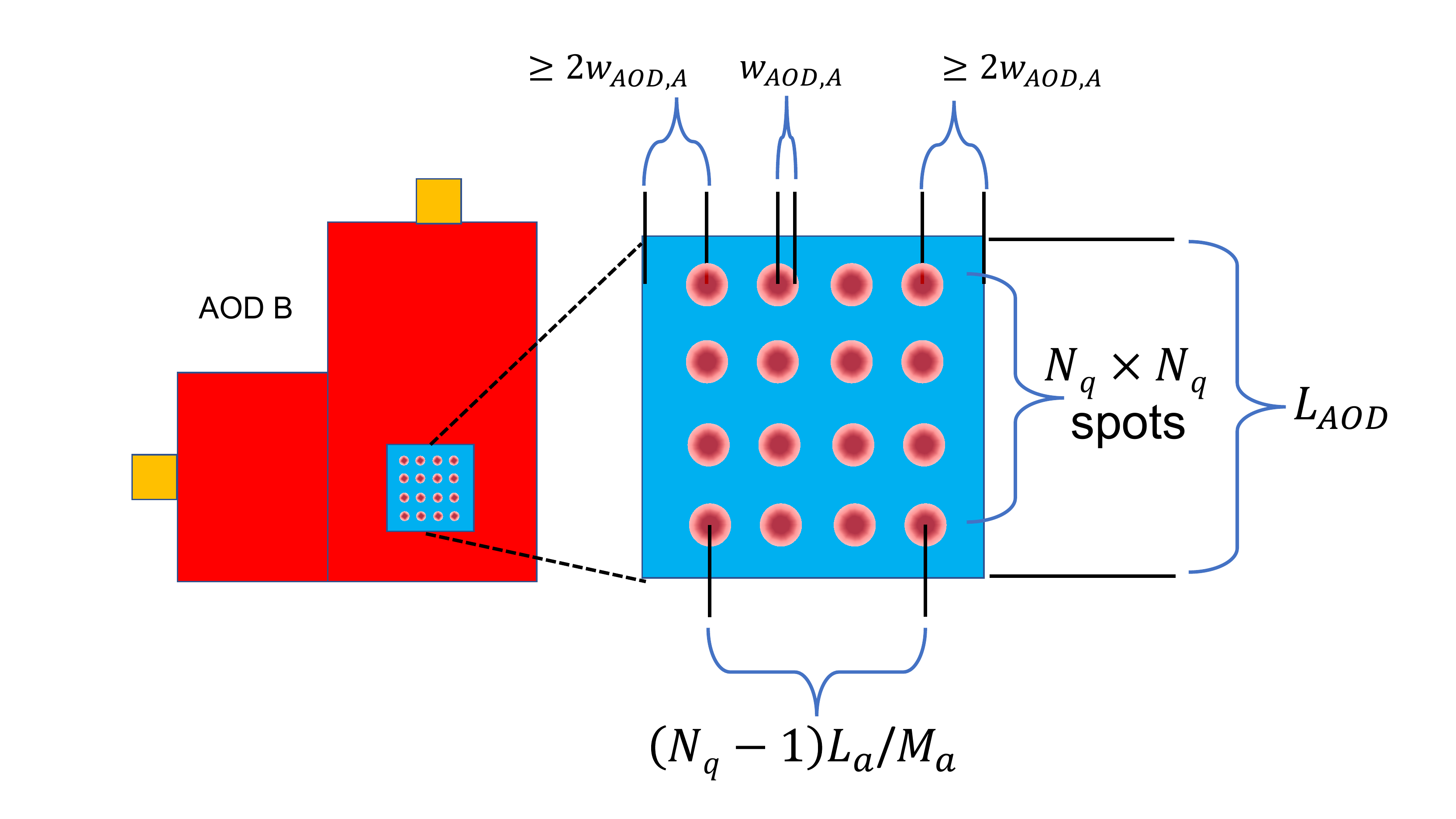}
\caption{Light beam parameters and dimensional constraints in acousto-optic deflector AOD B which is used  to correct the momentum shift that AOD A gives to the optical beam. Diffraction due to the SLM will cause the light field in AOD B to be different from AOD A, e.g the light image in AOD B might contain multiple beam spots. We assume that the SLM is addressing an array of $N_q \times  N_q$ qubits with addressing beam waists of $w_{\rm a}$ and qubit separations of  $L_{\rm a}$. If the waist of each Gaussian beam in AOD B is $w_{\rm AOD,A}$ (for simplicity, the waists in AOD A and B are taken to be equal), then the magnification from AOD B to the qubit plane is $M_{\rm a} = \frac{w_{\rm a}}{w_{\rm AOD,A}}$.  If a distance of $2 w_a$ is allowed between the AOD aperture edge and the nearest beam center to minimize clipping, then  in order to fit the beam image inside the active aperture of AOD B,  eq.  (\ref{eqn.AOD_B_clip}) must be satisfied.
\label{fig.AOD_B}
}
\end{figure}

For $q_{\rm SLM}=5$ and $q_{\rm a}=3$, we see that about $10 \times 10$ pixels are needed per site.  However, the diffraction efficiency is reduced for SLM patterns with a higher number of fringes over the SLM segment as shown in Sec. \ref{sec.simulations}. This decreased efficiency is due to the effective blazing of the SLM becoming less and less perfect as the fringe number approaches half the pixel number. Additionally, for higher diffraction angles, multi-site addressing without cross-talk, aberration correction, and beam shaping\cite{Gillen-Christandl2016} are error-prone and result in artifacts in the qubit array image plane.  To reduce these artifacts and  improve the diffraction efficiency over the addressable array, for $q_a= 3$ and $q_{\rm SLM}=5$, we conservatively allow $20 \times 20$ pixels per array site.  Note, that it is not necessary for all SLM segments to address the same set of array sites.  By applying an additional static or dynamic linear phase mask over each SLM segment, the portion of the qubit array that each segment addresses may be shifted. While this allows a larger array to be imaged, it reduces the number of times a particular site can be addressed before the SLM must be reset. 

For simplicity, we analyze the case where the entire array is addressed by each hologram patch.  The size of the array that can be addressed in this way is ultimately limited by the active aperture of AOD B.  This is because AOD B is in an image plane of the qubit array, so a magnified image of the addressing beams propagates through AOD B's aperture.  To be able to address all sites in the $N_q \times N_q$ array without clipping on the AOD aperture (here we set the Gaussian beam waists in AOD B and A  the same $(w_{\rm AOD,B}=w_{\rm AOD,A}$)
\begin{equation}
    L_{\rm{AOD}} \geq 4 w_{\rm AOD,A} +  (N_q-1)L_{\rm a}/M_{\rm a}   = 4 w_{\rm AOD,A} + q_{\rm a} w_{\rm AOD,A} (N_q-1),
\label{eqn.AOD_B_clip}
\end{equation}
where $M_{\rm a}$ is the magnification of the image from AOD B to the qubit array (see Fig. \ref{fig.AOD_B}). Equivalently, this can be written as
\begin{equation}
    1 \geq \frac{4+q_{\rm a}(N_q-1)}{q_{\rm AOD, A}}.
\end{equation}
Here we have assumed a square AOD aperture and allowed a $2 w_{\rm AOD,A}$ buffer between the edge of the AOD active aperture and the center of the nearest beam spot. 

\begin{table}[!t]
\caption{\label{tab.configuration1} Some possible examples of SLM, AOD, and beam size settings for configuration 1. For each example, the following quantities are listed in columns left to right: the total qubit array size $N_q\times N_q$, number of partitions into which the SLM is divided, the ratio of the beam  waist in AOD A (see Fig. \ref{fig.scanner_parameters}) to the AOD crystal length ($q_{\rm AOD, A}$), the average transition rate between different addressing patterns, and the burst rate between SLM resets.  Each of these settings is consistent with a $1000 \times 1000$ pixel SLM, 2D AODs with $11.5~ \mu$s transition time and time-bandwidth product  of TBW$=575$, SLM patch size to beam waist on the SLM ratio of $q_{\rm SLM}=5$, and array spacing to beam waist ratio of $q_a=3$.}
\centering
\begin{tabular}{ |c|c|c|c|c|  }
 \hline
 \multicolumn{5}{|c|}{Configuration 1 addressing examples} \\
 \hline
 qubit array & SLM partitions & $q_{\rm AOD, A}$ & average transition rate  & burst transition rate\\
 $N_q\times N_q$ & & &  $\left(10^3 ~\rm s^{-1}\right)$ &  $\left(10^3 ~\rm s^{-1}\right)$\\
\hline
  $7 \times 7$ & $7 \times 7$ & 52 & 39  &  181  \\

  $10 \times 10$ & $5 \times 5$ & 70 & 22  &  190  \\

  $25 \times 25$ & $2 \times 2$ & 180 & 4  &  198  \\
\hline
\end{tabular}
\end{table}

With these constraints, we observe that there is a trade-off between array size and number of partitions on the SLM. This translates to a trade-off between array size and the effective addressing speedup because the speedup of this configuration compared to using an SLM alone is related to the number of partitions the SLM is divided into. Assuming each SLM patch is used once before the SLM frame is reset, then the average time required to transition between SLM configurations is   
\begin{equation}
    T_{\rm ave} = \frac{1}{N_x N_y r_{\rm SLM}} + T_{\rm burst},
\end{equation}
where the SLM is divided into $N_x \times N_y$ patches, $r_{\rm SLM}$ is the full frame transition rate of the SLM, and $T_{\rm burst}$ is the time required for AODs A and B to transition between different patches on the SLM. Note that while the average scanner transition rate $1/T_{\rm ave}$ is limited primarily by the number of SLM partitions and the SLM refresh rate, AODs allow much faster transitioning between  segments on the SLM. The fast AOD transitions enable a burst rate which can be sustained until the SLM needs to be updated to address a new pattern of sites. Since multiple spots may be present in AOD B's aperture due to diffraction from the SLM, the time required for the acoustic wave to traverse the entire image region in AOD B is greater than or equal to the transition time of AOD A assuming the beam waist in AOD A is the same as the beam waist of the spots in AOD B. Allowing the acoustic wave time to travel at least $2 w_{\rm AOD,A}$ from the center of each beam spot, then an upper bound on the transition time ($T_{\rm{burst}}$) between these SLM partitions is
\begin{equation}
\label{eqn.burst_timeC1}
    T_{\rm{burst}} \leq \frac{4 w_{\rm AOD}+L_{\rm a} N_q}{v_{\rm a}}=\frac{T_{\rm{AOD}}(4 +q_{\rm a} N_q)}{q_{\rm AOD, A}},
\end{equation}
where $v_{\rm a}$ is the acoustic wave velocity in the AOD crystal. The burst rate is then $1/T_{\rm{burst}}$. Table \ref{tab.configuration1} lists some possible configurations assuming SLM and AOD parameters that   are currently commercially available.
This first configuration allows a significant speed-up over an SLM alone. Furthermore, because arbitrary addressing of the target regions can be achieved using each region of the SLM, gates can be operated in parallel over the qubit array.

\subsection{Configuration 2}
\label{conf.2}
The addition of a third 2D AOD (AOD C) to the scanner allows for greatly increased size of the addressable array and the average transition frequency, but comes at the expense of completely arbitrary addressing over the full array. This is made possible by setting the various SLM segments to address different combinations of sites within a small sub-array of sites using the same optical layout as described in configuration 1 of Fig. \ref{fig.configurations}. The addition of the final 2D AOD in configuration 2 of Fig. \ref{fig.configurations} allows this sub-array to be moved around a larger optical array. Note, as mentioned in the last section, it is possible to combine the roles of AODs A and B into a single 2D AOD (see Fig. \ref{fig.configurations_simplified}); however, as in the last section, we analyze  the configuration shown in Fig. \ref{fig.configurations} for simplicity.

\begin{figure}[!t]
\includegraphics[width=5in]{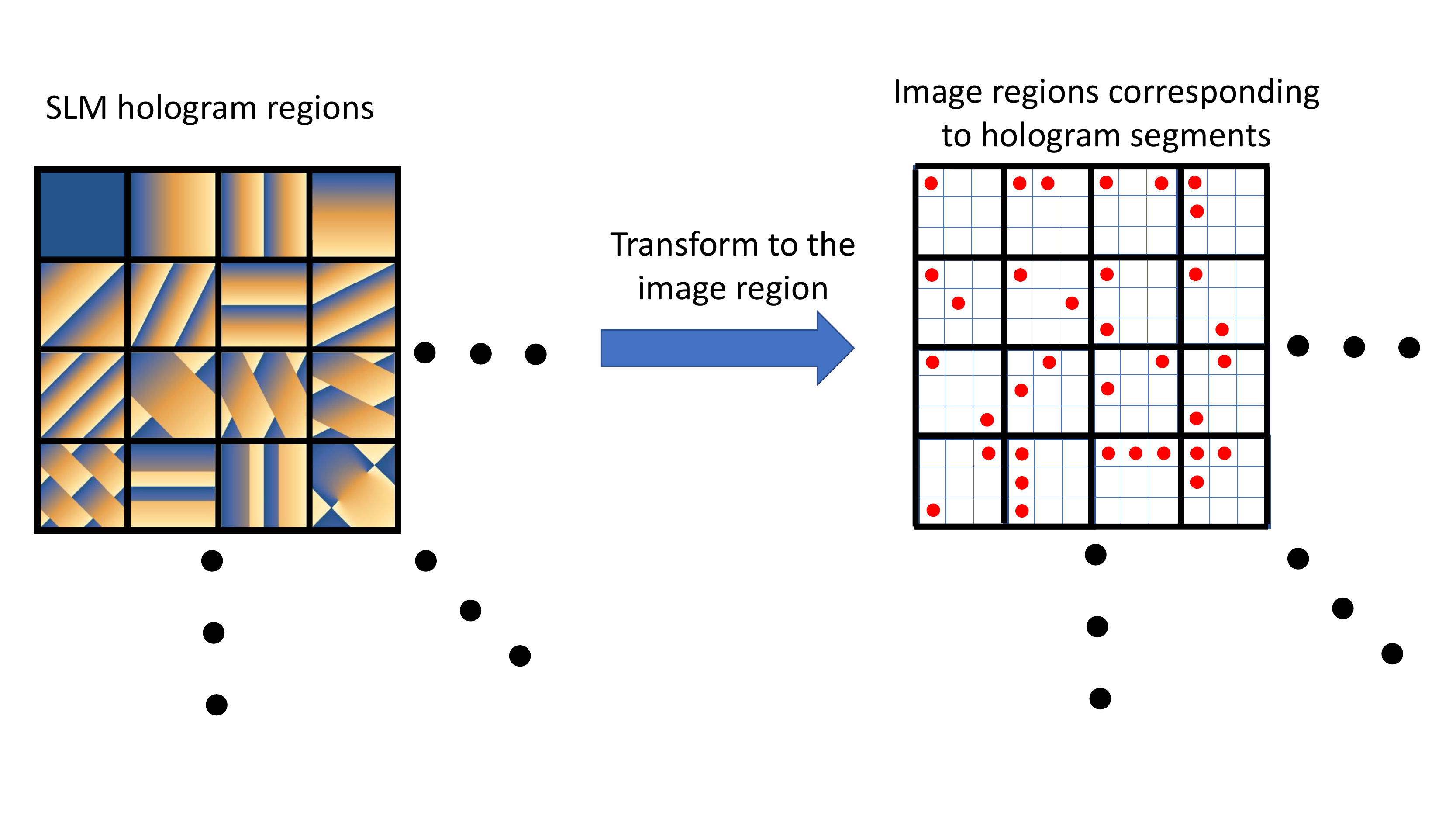}
\caption{
\label{fig.SLM_segmentation}An example of segmenting a SLM into multiple partitions for addressing a $3\times 3$ sub-array for use in configuration 2. In configuration 2, a third 2D acousto-optic deflector is used to position the SLM hologram images with respect to the qubit array. To address any single qubit in a $3 \times 3$ sub-array, only 1 hologram patch is needed.  Any two qubits in the sub-array can be addressed with the next 12 SLM configurations. Equation (\ref{eq.Pmax}) gives the number of SLM combinations required to simultaneously address $k$ qubits in an $m_s \times n_s$ sub-array.   
}
\end{figure}

Configuration 2 has many benefits. If holograms for all addressing combinations required for a quantum computation can be represented by a single SLM frame (see Fig. \ref{fig.SLM_segmentation}), then the SLM will not need to be reset during the computation.  The number of SLM patches needed to arbitrarily address $k$ qubits in a $m_s \times n_s$ sub array can be calculated from the combinatorial formula for putting $k$ identical particles in $m_s \times n_s$ boxes, $\frac{(m_s n_s)!}{k! (m_s n_s - k)!}$. However, since AOD C may be used to position the sub array, this formula over-counts the number of required configurations. Any two patterns which differ only by a translation can be represented by a single SLM patch and two AOD C settings. Without loss of generality, we quantify these patterns by counting only combinations which contain at least one spot in the left-most column and one spot in the top-most row, since any addressing pattern may be translated to satisfy this requirement. We can then subtract off combinations which do not meet the row criterion ($\frac{[(m_s-1) n_s]!}{k! [(m_s-1) n_s - k]!}$ combinations) and also the combinations where the column criterion is not met ($\frac{[m_s (n_s-1)]!}{k! [m_s (n_s-1) - k]!}$ combinations).  This subtraction double counts the combinations where both row and column criteria are not satisfied, so these ($\frac{[(m_s-1) (n_s-1)]!}{k! [(m_s-1) (n_s-1) - k]!}$ combinations) must be added back.  This results in the total number of SLM patches ($P$) needed to simultaneously address any combination of $k$ sites in a $m_s \times n_s$ sub array as
\begin{eqnarray}
\label{eq.Pmax}
    P(m_s,n_s,k) &=& \frac{(m_s n_s)!}{k! (m_s n_s - k)!} -\frac{[(m_s-1) n_s]!}{k! [(m_s-1) n_s - k]!} \nonumber\\
    &-&\frac{[m_s (n_s-1)]!}{k! [m_s (n_s-1) - k]!} + \frac{[(m_s-1) (n_s-1)]!}{k! [(m_s-1) (n_s-1) - k]!}.
\end{eqnarray}

\begin{table}[!ht]
\caption{\label{tab.sub_array_addressing} SLM patterns for different qubit addressing patterns in configuration 2. In configuration 2, the diffracted pattern from the SLM can be translated with respect to the array using AOD C. With this degree of freedom, the SLM only needs to address a sub-region to address the full qubit array. The number of SLM patches needed for arbitrary addressing of this sub-region depends on the dimensions of the sub-region ($m_s \times n_s$) and the maximum number of qubit sites to be simultaneously addressed, $k_{\rm max}$. The total number of SLM patches needed to arbitrarily address any combination of sites up to $k_{\rm max}$ sites simultaneously in the sub-region is then equal to $P_{\rm tot}(m_s,n_s,k_{\rm max})$. }
\centering
\begin{tabular}{| c|c|c|  }
 \hline
 \multicolumn{3}{|c|}{Configuration 2 SLM partitions} \\
 \hline
addressed qubit sub-region& simultaneously addressed sites&  unique configurations \\
$m_s \times n_s$ & $k_{\rm max}$ & $P_{\rm tot}(m_s,n_s,k_{\rm max})$\\
\hline
$3 \times 3$ & 2  & 13 \\

$3 \times 3$ & 3  & 61 \\

$3 \times 3$ & 4  &  158 \\

$4 \times 4$ & 2  &  25 \\

$4 \times 4$ & 3 &  229 \\

$5 \times 5$ & 2  &  41 \\

$5 \times 5$ & 3  &  621 \\
\hline
\end{tabular}
\end{table}

The number of partitions needed to address every sub-array combination for up to $k_{\rm max}$ simultaneous beams is then:
\begin{equation}
\label{eq.kmax}
    P_{\rm tot}(m_s,n_s,k_{\rm max}) = \sum_{k=1}^{k_{\rm max}}P(m_s,n_s,k).
\end{equation}  
Alternatively, one could include SLM patch optimization in the compilation step of the quantum computation.  This way, SLM patches which optimize gate parallelism in the quantum computation could be chosen.

In the case where SLM patches do not need to be reset during the calculation, the transition time between quantum gates will only be limited by the rise time of the three AODs, and the average transition rate will be equal to the burst transition rate.  Dimensions of the qubit array $N_q \times N_q$ that can be addressed can be much larger than the dimensions of the sub-array that is addressed by the SLM patches ($m_s \times n_s$ sites). The dimension $N_q$ is limited by the properties of AOD C, the light field entering the AOD, and the dimensions of the qubit array being addressed.  Assuming Gaussian inputs, then the size of the addressable array is
\begin{equation}
    \label{eqn.N_lC2}
    N_q = \frac{\pi}{q_{\rm AOD,C} q_{\rm a}}TBW,
\end{equation}
where $q_{\rm AOD,C}$ is the ratio of the AOD C's active aperture length to the beam waist in AOD C and both $q_{\rm a}$ and $TBW$ are defined in the same way as in the previous subsection (this analysis assumes that all three 2D AODs are the same model, though this need not be the case). Note that $q_{\rm AOD,C}$ can be very different from $q_{\rm AOD, A}$ since the corresponding beam waists are in conjugate Fourier planes. This difference results in the effective switching speed of the AODs being different.  The transition time needed for AOD B is larger than or equal to that of AOD A, so the transition time of the scanner is equal to the greater of the transition times of AOD B   and AOD C  

\begin{equation}
    \label{eqn.trans_timeC2}
    T_{\rm{transition}} =  \max \left( \frac{T_{\rm AOD,B}(4 +q_{\rm a} N_q)}{q_{\rm AOD, A}}, \frac{4 T_{\rm AOD,C}}{q_{\rm AOD,C}} \right).
\end{equation}

Depending on the size of the qubit array being addressed in each of the two configurations and the number of SLM partitions, the scanner transition time for configuration 2 might be slower than the burst transition time for configuration 1. However, the average transition time will typically be faster for configuration 2 since the SLM does not need to be reset.  Table \ref{Table.2} details the configuration 2 scanner performance given various scanner parameters. The values in this table were calculated for a $1000 \times 1000$ pixel SLM, however, higher resolution 4k ($4160 \times 2464$) SLMs are 
available commercially, and 8k ($8192 \times 4320$) pixel prototypes have been demonstrated \cite{Lazarev2012}.  These higher resolution SLMs would allow for correspondingly larger sub-array addressing and/or more complete addressing combinations.

\begin{table}[!ht]
\caption{Some possible examples of SLM and beam size settings for configuration 2. For each example, the following quantities are listed in columns left to right: the total qubit array size, number of partitions into which the SLM is divided, the dimensions of the sub-array that each SLM patch addresses, the maximum simultaneous addressing number $k_{\rm max}$ for which $P_{\rm tot}$ (see eq. (\ref{eq.kmax})) is less than or equal to the number of SLM partitions, the ratio of the beam  waist in AOD A (see Fig. \ref{fig.configurations}) to the AOD active aperture length ($q_{\rm AOD, A}$), the ratio of beam waist in AOD C to the active aperture length ($q_{\rm AOD,C}$), and the average transition rate between different addressing patterns. Note that in the last two rows, $k_{\rm max}$ is 2, but more than half of the 3 beam combinations could be included in the SLM partitions.  Each of these settings is consistent with a $1000 \times 1000$ pixel SLM, 2D AODs with $11.5 ~\mu$s transition time,   $TBW=575$, SLM patch size to beam waist on the SLM ratio $q_{\rm SLM}=5$, and a array spacing to beam waist ratio of  $q_{\rm a}=3$.}\label{Table.2}
\centering
\begin{tabular}{|c|c|c|c|c|c| c| }

 \hline
 \multicolumn{7}{|c|}{Configuration 2  addressing examples} \\
 \hline

    array size & SLM  & sub-array & $k_{\rm max}$ &$q_{\rm AOD, A}$ & $q_{\rm AOD,C}$ & transition rate\\
     $N_q\times N_q$ &partitions  & size &   &  & & $\left(10^3 ~\rm s^{-1}\right)$\\
\hline
$20 \times 20=400$ & $4 \times 4$ & $3 \times 3$ & 2 & 90 & 30 & 650  \\
 $40 \times 40=1600$ & $8 \times 8$ & $3 \times 3$ & 3 & 45 & 15 & 325  \\
 $67 \times 67=4489$ & $13 \times 13$ & $3 \times 3$ & 4 & 27 & 9 & 195  \\
 $30 \times 30=900$ & $5 \times 5$ &$4 \times 4$ & 2 & 72 & 20 & 416  \\
  $75 \times 75=5625$ & $12 \times 12$ & $4 \times 4$ & $2+$ & 30 & 8 & 173  \\
  $167 \times 167=27889$ & $12 \times 12$ & $4 \times 4$ & $2+$ & 30 & 3.6 & 78  \\
\hline
\end{tabular}
\end{table}

\begin{figure}[!t]
\includegraphics[width=5in]{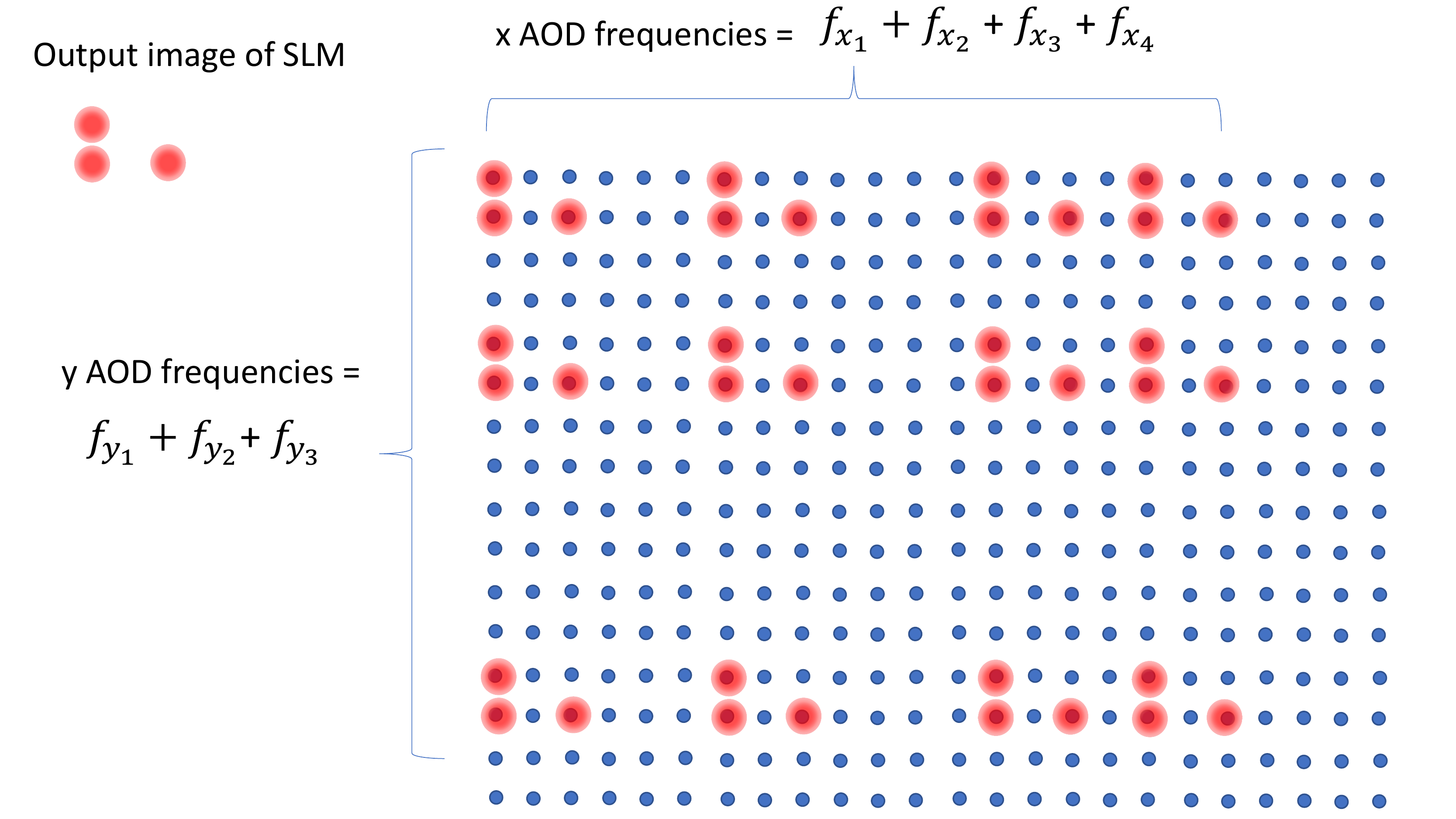}
\caption{
\label{fig.Multiple_AOD_Frequencies} By driving the final acousto-optic deflector (AOD) of configuration 2 (see Fig. \ref{fig.configurations}) with multiple frequencies, it is possible to duplicate an image in the qubit array region.  Each column (row) corresponds to a particular AOD frequency for the $x$ ($y$) axes of a 2D AOD.
}
\end{figure}

Since a single SLM only addresses a small section of the qubit array in configuration 2, it is not possible to perform completely arbitrary array addressing with configuration 2. However, it is possible to duplicate the image pattern created by the SLM in multiple regions of the qubit array by driving AOD C with multiple frequency tones (see Fig. \ref{fig.Multiple_AOD_Frequencies}). While this  multi-frequency driving does allow some addressing flexibility, it is limited because each axis of a 2D AOD can only provide momentum changes of an incident beam along the  corresponding axis. By driving that axis with multiple frequencies, an incident beam can be split since each acoustic wave frequency diffracts the beam by a different amount; however, the momentum shifts are all in the same direction. The resulting image will be duplicated by $N_{f_x} \times N_{f_y}$ times, where $N_{f_x}$ ($N_{f_y}$) corresponds to the number of $x$ ($y$) frequencies that the AOD axes are driven by.  

Furthermore, as mentioned above, the frequency of a beam diffracted by an acoustic wave is shifted by the acoustic wave frequency times the diffraction order. So the multiple images made by diffracting a beam with multiple acoustic wave frequencies will each receive a different frequency shift. When the resulting images are used, e.g., to drive atomic transitions, the various diffraction orders corresponding to the different acoustic wave frequencies will be detuned from the transition by different amounts. In some types of quantum gates, this relatively small frequency shift does not matter (e.g. ${\sf R}_z$ gates executed using differential Stark shifts on the qubit states \cite{Graham2019}, Raman gates with several GHz detuning \cite{Yavuz2006}, or gates using sub-picosecond laser pulses \cite{Song2018}).  Other quantum gates using resonant transitions might be affected by this detuning and will require additional addressing constraints or compensation strategies. For example, when driving two-photon transitions (each wavelength with its own scanner), then frequency shifts can be effectively compensated by having the frequency of the first and second photons of the transition being shifted by equal magnitudes but opposite signs \cite{Graham2022}.  Despite such limitations, the multi-tone operation is particularly useful for situations when identical operations need to be performed in several sub-regions over the array. Such situations arise  during initialization and syndrome measurement in logical qubits. Using multi-tone operation of AOD C, such operations can be performed in parallel, greatly saving operation time.

\section{Hologram Simulations}
\label{sec.simulations}

In the previous sections, we described the hybrid scanner in two different configurations. Here, we will show that these proposed configurations provide satisfactory performance even with the simplest hologram generation method which does not require iterative algorithms. The implementation of iterative algorithms might improve fidelity of the light field at the qubits at the cost of reduced efficiency and slower computation times. In general, when it comes to arbitrary beam shaping, it is necessary to have a large number of SLM pixels to precisely modulate the beam profile. Often, it also requires an advanced algorithm to gain high fidelity of the desired pattern, but it comes with the cost of losing power that should go to the target region. This is not desirable  because it will limit not only the diffraction efficiency but also sacrifice the number of different hologram patterns that can be simultaneously displayed in one frame. 

In the application to qubit addressing it is not always necessary  to impose the requirement of beam reshaping  on the holograms. In particular, when the shape of the input beam and the desired output beam are the same, or very similar, the hologram only has to deflect the input beam into single or multiple target sites within the target array.  Using this mechanism, the hologram can be easily generated by calculating the phase difference between the desired output field back-propagated to the SLM plane, $\mathcal{F}^{-1}\left(E_{\rm out}\right)$, and the input field in the SLM plane, $\mathcal{F}\left(E_{\rm in}\right)$. The required phase is 
\begin{equation}
    \phi = \text{mod [arg}[\mathcal{F}^{-1}\left(E_{\rm out}\right)-\mathcal{F}\left(E_{\rm in}\right), 2\pi]
\label{eq.phase}
\end{equation} 
where $E_{\rm in}$ is the input field,  $E_{\rm out}$, is the field of the desired output pattern, $ \mathcal{F} =$ Fourier Transform, and $\mathcal{F}^{-1}$  = inverse Fourier Transform.

\begin{figure}[!t] 
\centering
  \includegraphics[width=0.95\linewidth]{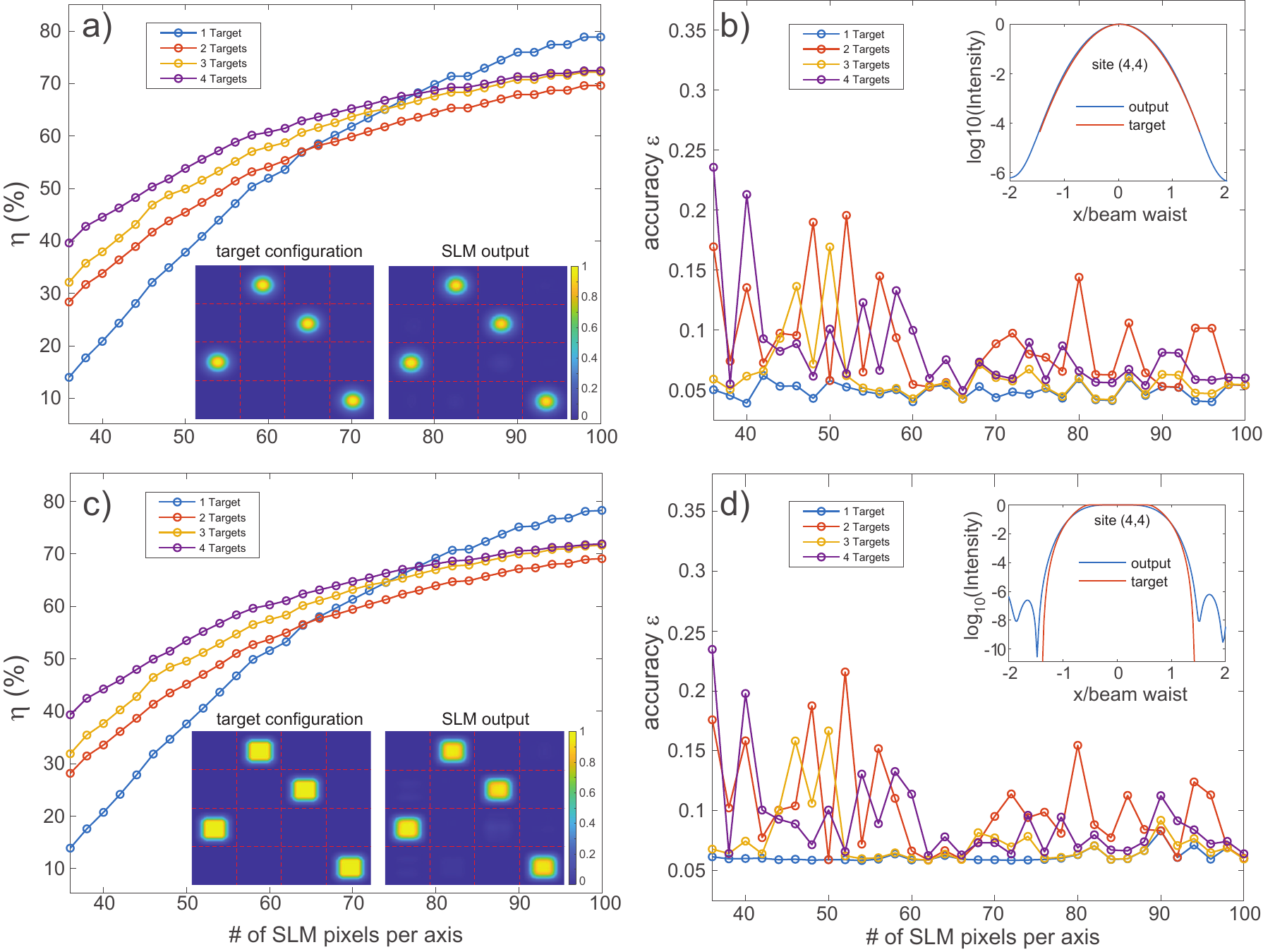}
\caption{Simulation results using Gaussian or square flattop beams as the target outputs $q_{SLM}=5$ and $q_{\rm a}=3$ to address up to 4 target locations in a $4\times 4$ array. a) Diffraction efficiency vs. number of SLM pixels per axis for Gaussian beams. c) Diffraction efficiency for flattop beams. b) Accuracy for Gaussian beams. d) Accuracy for flattop beams. The accuracy is the average for the number of targets used. The insets show the generated patterns as well as the  one-dimensional intensity profiles of the target and simulated output for the beam in the lower right corner of the array using 64 SLM pixels per axis with the target intensity outside the target region set to zero. }
\label{fig.simulation_results}
\end{figure}

Using this method we simulated the SLM performance for two different beam shapes: Gaussian  and square flattop beams as shown in  Fig. \ref{fig.simulation_results}. The flattop beam has  benefits over a Gaussian beam in qubit addressing because it not only reduces the crosstalk to neighboring sites but also the sensitivity of gate fidelity to optical misalignment\cite{Gillen-Christandl2010}. It can be incorporated in the  hybrid scanner architecture by using an additional flattop generator to shape the beam prior to passing through  AOD A. 

The simulations were done in Matlab based on Fourier analysis within the paraxial approximation, using a simulation array with $2^{13}\times 2^{13}$ pixels. Each SLM pixel is represented by 5 simulation pixels which allow other diffraction orders to show up in the output plane and therefore give an accurate value for the diffraction efficiency at the desired target region. The SLM is assumed to have 10 bits of phase resolution and the input beam illuminates the SLM normal to its surface, neglecting any angle shift. The finite fill factor of a real SLM was neglected when calculating diffraction efficiencies. In this simulation, We also assume aberration-free optics and a single polarization optical field, so the experimental results might slightly vary from the theoretical results presented in the paper. More detailed analysis and improvements to the results can be made by incorporating the full Helmholtz propagation.

In order to analyze the performance of the hybrid scanner, it is crucial to know how many pixels per SLM partition are needed to accurately address an $N_q\times N_q$  target array, while maintaining both the diffraction efficiency and the accuracy at a sufficient level. Using the simple method of eq. (\ref{eq.phase})  we analyze the number of SLM pixels needed for simultaneous  addressing of up to 4 target sites in a $4\times 4$ array.
To quantify the performance two metrics, diffraction efficiency $\eta$, and accuracy $\epsilon$, are used for the analysis and are defined as 
\begin{eqnarray}
  \eta &=& \frac{\text{total power in the target sites}}{\text{total input power on the SLM}}\\
      \epsilon &=& \sqrt{\frac{1}{N_{\chi}} \sum_{(x,y) \in \chi} \left(\frac{I_{\rm out}(x,y)-I_{\rm t}(x,y)}{I_{\rm t}(x,y)}\right)^2 }
\end{eqnarray}
where $\chi$ is the region covering the $4\times 4$ array of  target sites in the output plane  and $N_{\chi}$ is the number of points at which the output is sampled  \cite{Pasienski2008}. Here, the output intensity  $I_{\rm out}$ 
and target intensity $I_{\rm t}$ are normalized to have the same power in each target site. The points where $I_{\rm t}$ is smaller than $1/100$ of its maximum were omitted to avoid sensitivity to the edges of the beams when calculating $\epsilon$.
 
The simulations in Fig. \ref{fig.simulation_results} are relevant for a $1000\times 1000$ pixel SLM that is used in configuration 2 (see Sec. \ref{conf.2}). As shown in Table \ref{tab.sub_array_addressing} 
all possible configurations of simultaneously addressing up to 3 sites in a $4\times4$ sub-array requires 229 SLM partitions. Using an SLM with $1000\times 1000$ pixels each SLM partition can have $66\times66$ pixels. Using this number of pixels we achieve  $\eta\approx 60\%$ of total diffraction efficiency and accuracy $\epsilon\approx 0.05$. Less than $0.5\%$ of the total input power went to the untargeted sites within the target array and the rest of the missing power was distributed outside the target array. The results shown in the insets of Fig. \ref{fig.simulation_results} used 64 SLM pixels per axis to address a $4\times 4$ array which corresponds to $16\times 16$ pixels per array site in good agreement with the conservative estimate of $2N_{q, \rm max}\times 2 N_{q, \rm max}=20\times20$ pixels per array site as  discussed below  eq. (\ref{eq.Nqmax}).  Increasing the number of pixels to $80\times 80$ ($20\times 20$ per array site) we reach a diffraction efficiency of $\eta \approx70\%$ when targeting a single site, and a few percent lower when targeting up to 4 sites.  It is also possible to use more sophisticated hologram design methods such as Gerchberg-Saxton\cite{Gerchberg1972} or the MRAF algorithm of \cite{Pasienski2008}. These algorithms can achieve higher accuracy at the expense of lower diffraction efficiency. The results presented here verify the pixel requirements estimated in eq. (\ref{eq.Nqmax}), while ultimate performance limits will depend on the specific hologram design methodology employed.

\section{Conclusions}

In this paper, we described and analyzed the performance of two optical beam scanner architectures designed to provide fast quantum gate operations for future  generations of optically addressed qubit arrays. These scanners combine the arbitrary addressing of an SLM with the fast deflection capabilities of an AOD to provide site-selective quantum gates. The first of these configurations allows completely arbitrary site addressing of medium sized qubit arrays with a moderate transition rate between qubit gates. Completely arbitrary addressing allows parallel gates over the entire qubit array, but in this configuration there is a competition between average scanner transition rate (limited by the number of SLM partitions and the SLM transition rate) and the number of addressable sites in the array (limited by the number of pixels per SLM partition). Despite these trade-offs, we calculate a 100-site qubit array could be arbitrarily addressed with an average repetition rate of $22\times 10^3~\rm  s^{-1}$, a factor of 22 times speed-up over using an SLM alone. 

In the second configuration, the scanner is configured such that the SLM only provides arbitrary addressing on a sub-array of the qubit array and an additional AOD is added to allow scanning over the entire qubit array. In this second configuration, arbitrary addressing on the full qubit array is sacrificed for faster transition speeds and larger array sizes. With strategic SLM partitioning, this scanner enables fast addressing of large qubit arrays with arbitrary addressing on a movable sub-array within the qubit array. We show how to address a large array with more than 25,000 sites with limited multi-site simultaneous  addressing at a transition rate  of $78\times 10^3~\rm  s^{-1}$, a factor of 78 times the rate using an SLM alone.  Limited parallel addressing of the entire array could be achieved by driving the final AOD with multiple frequencies. Both of the scanner configurations can leverage the SLM for aberration correction and beam focus shaping to provide more optimal illumination patterns on targeted qubits, as is verified by numerical simulation of SLM phase holograms. These features combined will enable high fidelity, fast, and parallel gate operations in optically addressed qubit arrays. 

During revision of the paper we became aware of a patent \cite{DKim2021} related to the architecture presented here. The patent uses a fast switch to select different regions of a segmented SLM. This patent did not include $\bf k$-vector compensation in the atom plane nor the idea of scanning a sub-array relative to the full qubit array as is used in configuration 2.  

\section{Acknowledgement} This material is based on work supported by  NSF Award
2016136 for the QLCI center Hybrid Quantum Architectures and
Networks,
the U.S. Department of Energy Office of Science National Quantum
Information Science Research 
Centers, and   NSF Award 2210437.\\

\noindent
Disclosure: M. Saffman is a part time employee and equity holder in Infleqtion, Inc. . 

\bibliography{rydberg,optics,qc_refs,atomic,saffman_refs}

\end{document}